\documentclass{article}

\usepackage[english]{babel}
\usepackage[letterpaper,top=2cm,bottom=2cm,left=2cm,right=2cm,marginparwidth=1.75cm]{geometry}
\usepackage{amsmath}
\usepackage{graphicx}
\usepackage[colorlinks=true, allcolors=blue]{hyperref}
\usepackage{authblk}
\usepackage{appendix}
\usepackage{float}

\title{FDO Manager: Minimum Viable FAIR Digital Object Implementation}

\author[2]{Oussama Zoubia}
\author[1]{Nagaraj Bahubali Asundi}
\author[2]{Adamantios Koumpis}
\author[1,3]{Christoph Lange}
\author[2]{Sezin Dogan}
\author[1,2]{Oya Beyan}
\author[1,2]{Zeyd Boukhers}

\affil[1]{Fraunhofer Institute for Applied Information Technology FIT, Sankt Augustin, Germany}
\affil[2]{University Hospital of Cologne, Cologne, Germany}
\affil[3]{RWTH Aachen University, Aachen, Germany}
\date{}

\begin{document}
\maketitle

\begin{abstract}
In the digital age,  data has emerged as one of the most valuable assets across various sectors, including academia, industry, and healthcare.  Effective data preservation involves the management of data to ensure its long-term accessibility and usability.   Given  the importance  and  sensitivity  of  data,  the  need  for  effective  management is a crucial necessity.  One of the big recent proposed approaches for data management is the FAIR Digital Objects (FDOs) which has emerged to revolutionize the field of data management and preservation.  Central to this revolution is the alignment of FDOs with the FAIR principles (Findable, Accessible, Interoperable, Reusable), particularly emphasizing machine-actionability and interoperability across diverse data ecosystems.  This paper presents "FDO Manager" a Minimum Viable Implementation of FDOs, tailored specifically for the use case and field of research artefacts such as datasets, publications, and code. The paper discusses the core ideas behind the FDO Manager,  its architecture,  usage and implementation details,  as well as its potential impact, demonstrating a simple and abstract implementation of FDOs in the research realm.

\end{abstract}


\section{Introduction}

We live in a digital connected world,  where research plays an important role in our advancements and development in various fields.  The amount of published research papers have been increasing exponentially in the recent years and we currently live in a world full of Big Data, a term that describes the large amount of data; however, with this amount comes major challenges and concerns in its management and preparation for usability. It is reported that a large amount of that data is still difficult to find and to use; thus making a hurdle for researchers. The FAIR principles were proposed in the aim to solve the aforementioned issues and to promote a better data management and open science. Furthermore, the FAIR principles are mostly of theoretical nature and have very few adapted and implemented solutions, which serves as a motivation for our work to propose and implement a lightweight solution for the management of FDOs also called FDO Manager. The goal of our implementation makes the interaction between the user and FDOs simple and straightforward and also aligns with FAIR principles and FDO specifications. In the coming sections of this paper, related works and research will be explored and described, the proposed FDO Manager is then presented in detail in a later section; outlining its conceptual architecture and implementation details.

\newpage
\section{Related works}
Since its introduction in 2018~\cite{collins2018turning}, the concept of FAIR Digital Objects (FDOs) has seen a remarkable rise in interest within the research and data management communities. FDOs are key in enhancing data preservation and accessibility, contributing to a robust ecosystem for managing digital research resources~\cite{de2020fair}. This surge in popularity underscores the increasing relevance of the FAIR (Findable, Accessible, Interoperable, Reusable) Principles in the realm of scientific data management~\cite{wilkinson2016fair, mons2017cloudy}.

The FDO Framework~\cite{Bonino_da_Silva_Santos_2023} is built upon the FAIR Guiding Principles~\cite{wilkinson2016fair}, which are designed to enhance the FAIRness of digital objects,  focusing on concepts such as unique identification, rich metadata, etc.  However, implementing these specifications is challenging due to their purely theoretical nature. The recurring requirement of meta-data as FDOs, for instance, can lead to complex scenarios where the implementation becomes a task of balancing the depth of detail with practical usability.

The efforts of the European Open Science Cloud (EOSC) \footnote{\url{https://research-and-innovation.ec.europa.eu/strategy/strategy-2020-2024/our-digital-future/open-science/european-open-science-cloud-eosc_en}}
are closely aligned with ours, as it seeks to establish an infrastructure that enables researchers to utilize data efficiently, adhering to FAIR principles and fostering open access, Notably, EOSC-Life advances these goals by creating an open, digital, and collaborative space for biological and medical research.  Additionally, the EOSC-Life Workflow Collaboratory \cite{goble2023eosc} further facilitates  the  research  cycle  by  providing  an  accessible  ecosystem  for  researchers to find, use, and deploy complex workflows, thereby streamlining and enhancing the reproducibility of scientific processes.

RO-Crates \cite{RO-crate} are also closely related to our work,  as both focus on facilitating the management of research data in a structured format.  While RO-Crates provides a framework for packaging research outputs with their associated metadata, The FDO Manager similarly aims to simplify the management of FDOs and enhancing long term persistence through PIDs.

This paper contributes to the ongoing research on FDOs by presenting a simple and lightweight FDO implementation which will be further explained in later sections. This implementation can be used for managing FDOs and can serve as a foundation and could be integrated with a more advanced implementations.

\section{FDO Manager}

The core concept driving the FDO Manager is to present a straightforward and minimalist solution that not only conforms to the FAIR Principles but also addresses the practical challenges of managing complex digital objects and their metadata in a systematic and user-centric manner.  

\begin{figure}[H]
    \centering
    \includegraphics[width=0.7\textwidth]{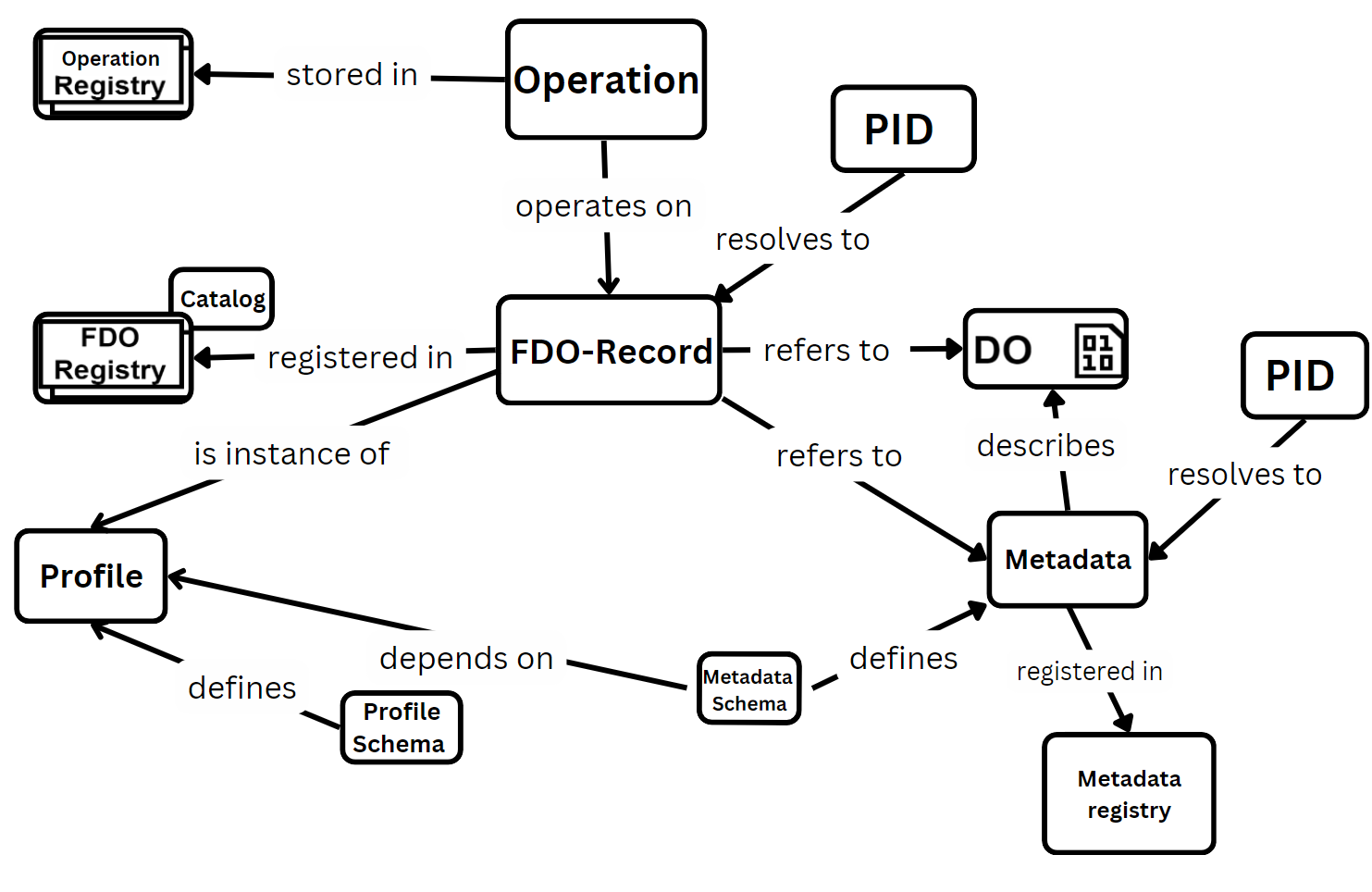}
    \caption{The FDO Manager architecture}
    \label{FDO:Schema}
    
\end{figure}

Figure~\ref{FDO:Schema} presents a refined view of the FDO Manager’s conceptual architecture, which outlines the various components and their interactions in the ecosystem. Central to this structure is the FDO Record, which refers to a digital object.  The metadata of the digital object is separated from the object itself, in alignment with the FDO specifications.  This separation ensures that the metadata persists even if the digital objector its record is lost or deleted.  Each type of FDO can have a specific operation applied to it, thus, the operations component is aimed to solve that and to also enhance interoperability by specifying the various operations that can be applied to an FDO. In the architecture, PIDs (Persistent Identifiers) are used to ensure the persistency and findability of both the FDO record and the metadata. 

The architecture also includes an FDO Registry, Metadata Registry, and Operations Registry, which serve as indexable catalogs within the ecosystem. Since the aim is to store and manage digital objects, having search and index functionality is essential to facilitate the findability within the FDO Manager.

Profiles are primarily used to specify the different types of FDO records that the FDO Manager handles.  For instance, in the realm of research artifacts, these profiles might include datasets, code, and publications. 

In the pursuit of interoperability, our
approach leverages the widely recognized Schema.org\footnote{\url{https://schema.org/}} standards to define the metadata schema for the proposed FDO records. which ensures a standardized and universally understood structure for metadata, enhancing the compatibility and exchange of information within the broader digital ecosystem.

\begin{figure}[H]
    \centering
    \includegraphics[width=0.4\textwidth]{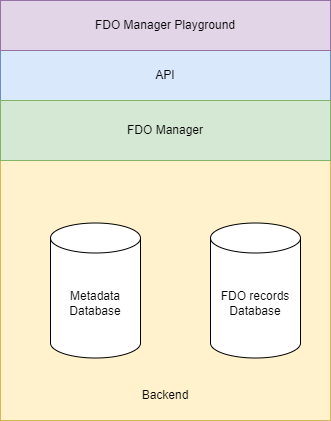}
    \caption{The FDO Manager layers}
    \label{FDO:Layers}
    
\end{figure}

Figure \ref{FDO:Layers} illustrates a layered architecture of the FDO Manager, which illustrates the different levels of abstraction of the functionalities and indicating which layers are accessible to users. It mainly consists of four main layers: 
\begin{itemize}
    \item \textbf{Backend layer:} This layer defines the databases for both metadata and FDO records.
    \item \textbf{FDO Manager Layer:} This core part of the codebase contains the primary logic and functionality of the FDO Manager.
    \item \textbf{API layer:} The FDO Manager is exposed to users through a set of API endpoints, allowing them to manage metadata and FDO records.
    \item \textbf{FDO Manager Playground:} an additional frontend layer provides a user-friendly interface for interacting with the FDO Manager and testing its functionalities.
\end{itemize}

From a user perspective only the two first layers are accessible to the users which are  the  API  and  the  FDO  Manager  Playground,  this  abstraction  and  design  choice simplifies the interaction ensuring that users can effectively use the FDO Manager.  In practice,  users first input metadata into the FDO Manager,  then proceed to register FDO  records  associated  with  this  metadata.   This  sequence  facilitates  an  efficient, structured approach to managing and accessing digital objects.

For seamless integration with other services and implementations, the API can be utilized, providing all the necessary endpoints for interacting with the FDO Manager.

To explore how metadata and FDO records are structured and related within the FDO Manager, we turn to Figure 3 which presents two tables:  one containing information about the FDO record and the other detailing the associated metadata.  The relationship between these tables is established through a reference metadata PID, ensuring data integrity and coherence between the FDO records and their corresponding metadata.

For further details about the metadata of the profiles, please refer to our comprehensive  documentation  available  on  our website\footnote{\url{https://fdda1.gitlab.io/fdom/schemas/overview/}}. This  documentation  provides  an in-depth overview of the schema definitions for the profiles, their description and their expected type; as well as their relationship to schema.org classes. The documentation resources are continuously updated to reflect any changes or enhancements, ensuring that the information remains current and accurate. 

\begin{figure}[H]
    \centering
    \includegraphics[width=0.8\textwidth]{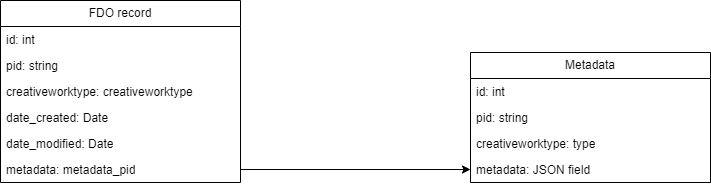}
    \caption{The FDO Manager databases architecture}
    \label{FDO:Databases}
    
\end{figure}

\section{Conclusion}
In conclusion, the FDO Manager emerges as a practical and lightweight solution for FDO  management.   Explored  in  this  paper  are  its  architecture,  implementation  de-tails,  and  alignment  with  FDO  specifications  and  FAIR  Principles.   With  a  focus  on research  artifacts,  the  FDO  Manager streamlines  workflows,  enhances  persistence, and enforces discoverability, contributing significantly to open science. Its adoption ensures that research artifacts remain easily accessible, making a valuable contribution to transparent and collaborative scientific practices. Certain limitations are anticipated, scalability remains an open question, specially under high data or user loads, which would require empirical testing to validate.

\section*{Author contributions}
\textbf{O.Z:} Writing -- Original Draft Preparation, Methodology, Software, Visualization, \textbf{Z.B:} Writing -- Review \& Editing, Conceptualization, Methodology, Supervision, Project Administration \textbf{N.A:} Conceptualization, Methodology, \textbf{S.D:} Visualization, \textbf{A.K:} Conceptualization, Supervision, Project Administration, \textbf{C.L:} Writing -- Review, Funding Acquisition, \textbf{O.B:} Conceptualization, Supervision, Project Administration, Funding Acquisition.

\section*{Competing Interests}
The authors declare that they have no competing interests.

\bibliographystyle{plain}

\begin{thebibliography}{1}

\bibitem{Bonino_da_Silva_Santos_2023}
Luiz~Olavo Bonino~da Silva~Santos, Tiago~Prince Sales, Claudenir~M. Fonseca, and Giancarlo Guizzardi.
\newblock {\em Towards a Conceptual Model for the FAIR Digital Object Framework}.
\newblock IOS Press, December 2023.

\bibitem{collins2018turning}
Sandra Collins, Francoise Genova, Natalie Harrower, Simon Hodson, Sarah Jones, Leif Laaksonen, Daniel Mietchen, R{\=u}ta Petrauskait{\.e}, and Peter Wittenburg.
\newblock Turning {FAIR} into reality: Final report and action plan from the european commission expert group on {FAIR} data.
\newblock 2018.

\bibitem{de2020fair}
Koenraad De~Smedt, Dimitris Koureas, and Peter Wittenburg.
\newblock {FAIR} digital objects for science: From data pieces to actionable knowledge units.
\newblock {\em Publications}, 8(2):21, 2020.

\bibitem{goble2023eosc}
Carole Goble, Fiona Bacall, Stian Soiland-Reyes, Stuart Owen, Ines Eguinoa, Benoit Droesbeke, and Frederik Coppens.
\newblock The eosc-life workflow collaboratory for the life sciences.
\newblock In {\em Proceedings of the Conference on Research Data Infrastructure}, volume~1, 2023.

\bibitem{mons2017cloudy}
Barend Mons, Cameron Neylon, Jan Velterop, Michel Dumontier, Luiz Olavo~Bonino da~Silva~Santos, and Mark~D Wilkinson.
\newblock Cloudy, increasingly {FAIR}; revisiting the {FAIR} data guiding principles for the european open science cloud.
\newblock {\em Information services \& use}, 37(1):49--56, 2017.

\bibitem{RO-crate}
Stian Soiland-Reyes et~al.
\newblock Packaging research artefacts with ro-crate.
\newblock {\em Data Science}, 1:97--138, January 2022.

\bibitem{wilkinson2016fair}
Mark~D Wilkinson, Michel Dumontier, IJsbrand~Jan Aalbersberg, Gabrielle Appleton, Myles Axton, Arie Baak, Niklas Blomberg, Jan-Willem Boiten, Luiz~Bonino da~Silva~Santos, Philip~E Bourne, et~al.
\newblock The {FAIR} guiding principles for scientific data management and stewardship.
\newblock {\em Scientific data}, 3(1):1--9, 2016.

\end{thebibliography}

\end{document}